\documentclass[
    ,final            
  ]
  {aipproc}

\layoutstyle{6x9}

\usepackage[T1]{fontenc}
\usepackage[latin9]{inputenc}
\usepackage{amsmath}
\usepackage{amsfonts}
\usepackage{amssymb}
\usepackage{latexsym}

\usepackage{graphicx}
\usepackage{pst-node,lscape,amsfonts,amsbsy,graphics,lscape,mathrsfs,dsfont,color}
\newcommand{\bea}{\begin{eqnarray}} \newcommand{\eea}{\end{eqnarray}}

\newcommand{\sst}{\scriptscriptstyle}
\newcommand{\hI}{\widehat{I}}

\begin{document}

\title{Final State Interactions, T-odd PDFs \& the  Lensing Function}


\author{Leonard Gamberg}{
  address={Penn State University Berks, Department of Physics,
Reading,Pennsylvania 19610-USA
}}

\author{Marc Schlegel}{
  address={Theory Center, Jefferson Lab, 
Newport News, Virginia 23606-USA
}}

\begin{abstract}
It has been suggested 
that under certain  approximations 
the Sivers effect
can be described in terms of 
factorization of final state 
interactions and a spatial distortion of impact parameter 
space parton distribution; that is 
a convolution of the so-called lensing function and 
the impact parameter GPD $E$. In this approach the lensing function 
is calculated in a non-perturbative eikonal model. This enables a 
comparison between the a priori distinct Sivers function and the 
GPD $E$ which goes beyond the discussion of overall signs.
\end{abstract}

\maketitle


``T-odd''
transverse momentum-dependent (TMD) parton distribution 
functions (PDFs)
have gained considerable attention in recent years. 
Prominent examples, and  the subject of
extensive
study, are the Sivers and Boer-Mulders functions~\cite{Sivers:1989cc,Boer:1997nt}.
In the factorized picture of 
semi-inclusive deep inelastic scattering~\cite{Mulders:1995dh,Ji:2004wu}
at small transverse momenta 
$P_T\sim k_\perp << \sqrt{Q^2}$
the Sivers effect describes a 
transverse target spin asymmetry (TSSA)   
through the ``naive'' T-odd structure,
$\Delta f(x,\vec{k}_T)\sim
 S_T\cdot(\widehat{P}\times\vec{k}_T) f_{1T}^\perp(x,k_T^2)$.
Dynamically, ``T-odd''-PDFs  emerge from the gauge link structure of the 
multi-parton quark and/or gluon
correlation functions~\cite{Collins:2002kn,Belitsky:2002sm,Boer:2003cm}
which describe initial/final-state interactions (ISI/FSI)
of the active parton via soft gluon exchanges with the target remnants.
Though these interactions  are 
non-perturbative recent 
phenomenological  calculations 
approximate 
the FSIs by  perturbative one-gluon exchange~\cite{Brodsky:2002cx,Ji:2002aa,
Goldstein:2002vv,Gamberg:2003ey,Bacchetta:2003rz,Lu:2004au,Gamberg:2007wm,
Bacchetta:2008af}.  
We improve  this approximation~\cite{gambschl:2009gs} by 
applying non-perturbative eikonal methods~\cite{Abarbanel:1969ek,Fried:2000hj} 
to calculate
higher-order soft gluon contributions.
This approach provides  insight into
the  picture of the Sivers effect
 proposed by Burkardt in the 
mixed phase space  
$\{x,\vec{b}_T\}$ description of the 
nucleon~\cite{Burkardt:2002ks,Burkardt:2003uw}.
We consider
the first ${k}_T$  moment of the Sivers
function
written
as a gluonic pole matrix element
in impact parameter space~\cite{Meissner:2007rx}
\bea
\langle k^{i}_T(x)\rangle\equiv
\int d^{2}b_{T}\int \frac{dz^{-}}{4\pi}\mathrm{e}^{ixP^{+}z^{-}}
\langle{ P^{+},\vec{0}_{T}; S_{T}}|\bar{q}(z_1)\gamma^{+}
 {[z_{1}; z_{2}]}I^{i}(z_{2}) q(z_{2})|{
P^{+},\vec{0}_{T}; S_{T}}
\rangle
\label{eq:GPME}
\eea
where $\langle k^i_{\sst T}(x)\rangle=M\epsilon_{\sst T}^{ji}S_{\sst T}^{j}
f_{\sst 1T}^{\perp(1)}(x)$.   The 
operator $\widehat{I}$ originates from the time-reversal
behavior of the FSIs  implemented by the
gauge link operator  $[a\,;\, b]$ which denotes 
a  straight
Wilson line connecting the two locations $a$ and $b$.  
It is a function of the gauge-link and  the gluonic field strength 
tensor, $I^{i}(\tfrac{z^{^{-}}}{2}n)=\int dy^{-}\,[\tfrac{z^{^{-}}}{2}n\,,\, y^{-}n]\, gF^{+i}(y^{-}n)\,[y^{-}n\,,\,\tfrac{z^{^{-}}}{2}n]$  where 
 $n^{\mu}=1/{\sqrt{2}}(1,0,0,-1)$ is the light-like vector, 
and $z_{1/2}=(\mp\tfrac{z^{-}}{2\sqrt{2}},\vec{b}_{T},\pm\tfrac{z^{-}}{2\sqrt{2}})$.  We find that 
calculating the operator  $\widehat{I}$ 
in an eikonal approximation, 
including multiple gluon exchanges, 
the integrand of
Eq.~(\ref{eq:GPME}) 
factorizes into a distortion of the
transverse space parton distribution and the FSIs
resulting in the relation~\cite{Burkardt:2003uw}
\begin{equation}
\displaystyle
-M^2\epsilon_{T}^{ij}S_{T}^{j}f_{1T}^{\perp (1)}(x)\simeq\int d^{2}b_{T}\,\mathcal{I}^{i}(x,\vec{b}_{T})\epsilon_{T}^{ij}b_{T}^{i}S_{T}^{j}\,
\partial_{\scriptscriptstyle \vec{b}_T^2}(\mathcal{E}(x,\vec{b}_{T}^{2})).
\label{eq:Relation}
\end{equation}
Here  $\mathcal{I}$ represents 
the ``chromodynamic lensing function''.
While such a factorization 
doesn't hold in general~\cite{Meissner:2008ay,Meissner:2009ww}  it is
 fulfilled in lowest order contributions of simple field-theoretical
models \cite{Burkardt:2003je,Meissner:2007rx}. 
In particular, contributions
beyond the lowest order, where the quark fields and the operator $\hI$
``interact'' 
(they are interacting Heisenberg operators) 
leads to a breakdown of the relation which we write as 
$I^{i}=\mathcal{I}^{i}(x,\vec{b}_{T})\mathds{1}+\mathrm{b. t.}$, 
where $\mathrm{b.\, t.}$ denote the  ``breaking terms''. 
We estimate  the size of these breaking terms by deriving the lensing
function in a relativistic eikonal model~\cite{gambschl:2009gs}. 

\begin{figure}
\includegraphics[scale=0.56]{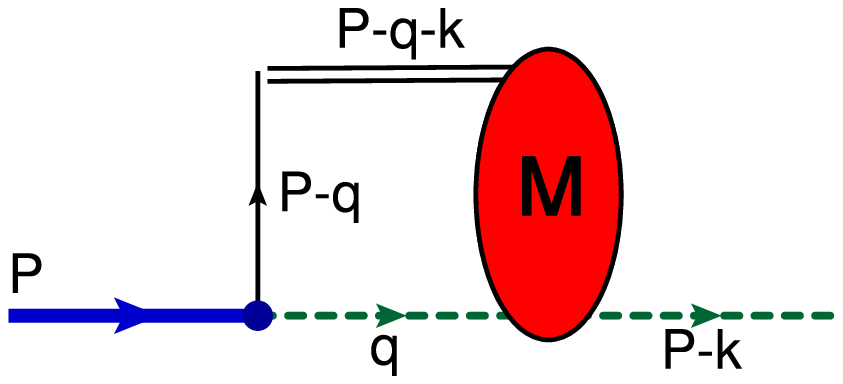}~~~~~~~~~
\includegraphics[scale=0.56]{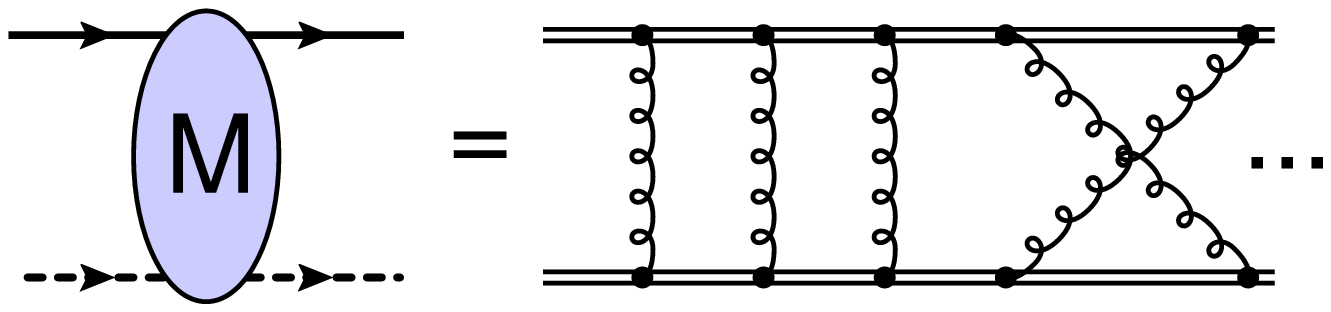}~~~
\caption{\label{fig:TMD-amplitude-including}
Left: TMD amplitude $W(P,k;S)$ including FSIs for re-scattered
eikonalized quark and the spectator. The final state interactions
are described by a non-perturbative amputated 
scattering amplitude $\mathsf{M}$. Right: FSI in the relativistic eikonal
model where quark and spectator are eikonalized. Only ladder diagrams 
are taken into account.}
\end{figure}

\begin{figure}
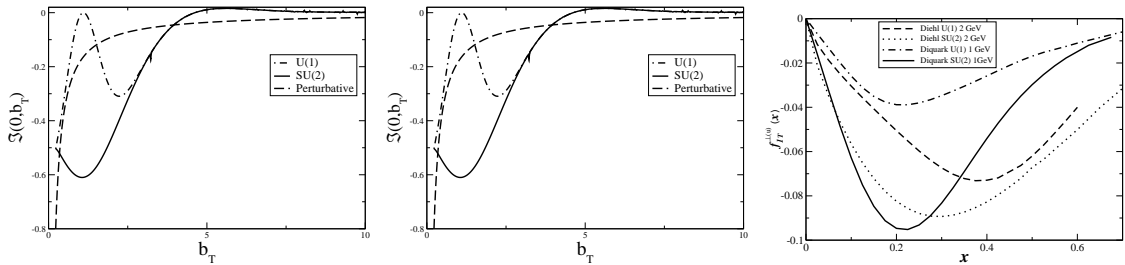

\includegraphics[scale=0.195]{LensFunc_new2.eps}~
\includegraphics[scale=0.195]{EDT_new2.eps}~
\includegraphics[scale=0.195]{Sivers_new2.eps}
\caption{\label{fig:Lens-GPD-Sivers}
Left: Lensing function for the perturbative U(1) and non-Abelian
SU(2) cases. Note that a value of $\alpha=0.3$ in the perturbative result
is arbitrarily chosen. Center: Valence u-quark GPD 
$E^{(u)}(x,0,-\Delta_T^2)$ at $x=0.3$ in spectator model~\cite{gambschl:2009gs} and from the parameterization~\cite{Diehl:2004cx}. Right: First 
moment of the u-quark Sivers function from Eq.~(\ref{eq:Relation}) 
for the  two models~\cite{Diehl:2004cx,gambschl:2009gs} of input GPD.}
\end{figure}

If the FSIs are modeled by a one-gluon exchange, the lensing function
can be unambiguously identified in a diquark
spectator model~\cite{Burkardt:2003je}. For this reason we choose
to stay in the diquark picture but describe the FSIs by a non-perturbative
(re-)scattering  at the amplitude level as in
Fig.~(\ref{fig:TMD-amplitude-including}). 
We perform a calculation of $\mathsf{M}$ and in turn the lensing function
in   a $U(1)$ theory, and a  $SU(2)$-non-Abelian theory in an attempt to
study the effects of color on  the relation Eq.~(\ref{eq:Relation}).
Thus, we do not take into account gluon 
exchanges  
 between the quark line and the soft blob
since they give rise to interactions between the
quark fields and the operator $\hI$ in Eq.~(\ref{eq:GPME}) which 
 lead to a breakdown of the relation, Eq.~(\ref{eq:Relation}). 
 Self-interactions and real gluon emission 
between quark and diquark lines 
are accounted for through masses and 
vertex form factors in the spectator model.
Applying the usual Feynman rules, 
and expressing
the FSIs through the amputated quark-diquark scattering
amplitude $\mathsf{M}$ leads to the expression
\begin{eqnarray}
\Delta W_{i}(P,k;S)
=
 \hspace{-.15cm}\int\frac{d^{4}q}{(2\pi)^{4}} \frac{ig_{N}\left((P-q)^{2}\right)
\left[(\slash{\hspace{-.25cm} P}-\slash{\hspace{-.25cm}q}+m_{q})u(P,S)\right]_{i}\mathsf{M}(q,P-k)}{\left[n\cdot(P-k-q)+i0\right]\left[(P-q)^{2}-m_{q}^{2}+i0\right]\left[q^{2}-m_{s}^{2}+i0\right]}+\mathrm{b.t.}, 
\label{eq:StartW}\end{eqnarray}
where $i(n\cdot(P-k-q)+i0)^{-1}$
represents the
eikonal propagator, the double-line in Fig.~\ref{fig:TMD-amplitude-including},  and  $\Delta W=W-W^0$ where $W^0$ 
 denotes the contribution
without final-state interactions. 
Performing 
the contour-integration on $q^{-}$
we pickup  poles only from the denominators in Eq.~(\ref{eq:StartW})
where we assume that  $\mathsf{M}$ does not
contain poles in $q^{-}$.
This is not generally
true even in the one-gluon exchange approximation~\cite{Gamberg:2006ru,Gamberg:2007wm} 
where poles exist for  higher twist T-odd TMDs
for a scalar spectator and also at leading twist  for an axial
vector spectator. However, these contributions  
give rise to light-cone divergences~\cite{Gamberg:2006ru}
which would lead to a breakdown of Eq.~(\ref{eq:Relation}).
Next considering the  integration on $q^{^{+}}$
the eikonal propagator can be split into a real and imaginary part.
While the imaginary part of the eikonal propagator forces the
diquark to be on-shell and restores the relation, Eq.~(\ref{eq:Relation})
the real part leads to contributions beyond the relation and has
to be attributed to the breaking terms.
The resulting kinematics are consistent with the picture that
quark and diquark  move quasi-collinearly with respect to the nucleon
and that
 the FSIs are dominated by ``small''
transverse momenta gluon exchanges.
Now as  input to Eq.~(\ref{eq:Relation}) 
we calculate the GPD $E(x,0,-\vec{\Delta}^2)$
for a scalar diquark spectator with a multi-pole 
form factor~\cite{gambschl:2009gs}
\begin{eqnarray}
E(x,0,-\vec{\Delta}_{T}^{2})=  
\frac{g^{2}(1-x)^{2}}{(2\pi)^{3}}\int {\scriptstyle{d^{2}k_{T}
\tfrac{\left((1-x){\Lambda}^2\right)^{2n-2}M(xM+m_{q})}{\left[(\vec{k}_{T}-\tfrac{1}{2}(1-x)\vec{\Delta}_{T})^{2}+\tilde{\Lambda}^{2}\right]^{n}\left[(\vec{k}_{T}+\tfrac{1}{2}(1-x)\vec{\Delta}_{T})^{2}+\tilde{\Lambda}^{2}\right]^{n}}}}
, \label{eq:E}
\end{eqnarray} 
where $\Lambda$, $m_q$, $m_s$ and $M$ are ``cutoff'', 
quark, spectator, and nucleon masses
respectively and 
$\tilde{\Lambda}^2=xm_s^2-x(1-x)M^2+(1-x)\Lambda^2$.
We  transform to impact parameter space,
$\mathcal{E}(x,\vec{b}_{T}^{2})=\int\frac{d^{2}\Delta_{T}}{(2\pi)^{2}}\,\mathrm{e}^{-i\vec{\Delta}_{T}\cdot\vec{b}_{T}}E(x,0,-\vec{\Delta}_{T}^{2})$ and
we express the lensing function $\mathcal{I}$  in terms of the
real and imaginary part of the 
scattering amplitude $\mathsf{M}$,\begin{equation}
\mathcal{I}^{i}(x,\vec{q})=
\int\frac{d^{2}p}{(2\pi)^{2}}\,(2\vec{p}-\vec{q})^{i}\,
\frac{\Im[\mathsf{M}](x,|\vec{p}|)}{4(1-x)P^{+}}\left((2\pi)^{2}\delta^{(2)}(\vec{p}-\vec{q})+\frac{\Re[\mathsf{M}](x,|\vec{p}-\vec{q}|)}{4(1-x)P^{+}}\right),\label{eq:LensFunc}\end{equation}
where $\vec{p}$ and $\vec{q}$ are transverse momentum
integration variables,
and calculate $\mathsf{M}$ 
in the eikonal approximation
(details can
be found in~\cite{gambschl:2009gs})
\begin{equation}
\mathsf{M}^{\mathrm{eik}}(x,|\vec{q}|)=4\pi(1-x)P^{+}\int_{0}^{\infty}dz\, z\, J_{0}(z|\vec{q}|)\left(\mathrm{e}^{i\chi(z)}-1\right), 
\label{eq:eikonalAmplitude}
\end{equation}
where $\chi(z)$ is the eikonal phase in configuration space
\begin{eqnarray}
\chi(|\vec{z}_{T}|) & \equiv & -e_{q}e_{s}n^{\rho}\bar{n}^{\sigma}\int_{-\infty}^{\infty}d\alpha\int_{-\infty}^{\infty}d\beta\mathcal{D}_{\rho\sigma}^{-1}(z+\alpha n-\beta\bar{n})
, \label{eq:Phase}
\end{eqnarray}
$\bar{n}=1/{\sqrt{2}}(1,0,0,1)$, 
 $e_{q}e_{s}=-4\pi\alpha$
and $\mathcal{D}_{\rho\sigma}^{-1}(z)$ are the 
quark-gluon coupling and gluon propagator respectively.  
Eq.~(\ref{eq:eikonalAmplitude}) is valid in an U(1) Abelian gauge theory. 
Using Eq.~(\ref{eq:eikonalAmplitude})
for $\mathsf{M}^{\mathrm{eik}}$ and
transforming to  impact parameter space the lensing function 
\begin{equation}
\mathcal{I}_{\rm{\scriptscriptstyle{U(1)}}}^{\mathrm{eik},i}(x,\vec{b}_{T})=\frac{b_{T}^{i}}{4|\vec{b}_{T}|}\chi^{\prime}(\tfrac{|\vec{b}_{T}|}{1-x})\left(1+\cos\chi(\tfrac{|\vec{b}_{T}|}{1-x})\right). 
\label{eq:eikonalLensFunctionTa}
\end{equation}
An attempt to build in color into the eikonal formalism has been provided
in~\cite{Fried:2000hj}. Using these results  
the lensing function in an SU(2) gauge theory becomes
\begin{equation}
\mathcal{I}_{\rm{\scriptscriptstyle{SU(2)}}}^{\mathrm{eik},i}(x,\vec{b}_{T})
=\frac{b_{T}^{i}}{4|\vec{b}_{T}|}
\chi^{\prime}(\tfrac{|\vec{b}_{T}|}{1-x})
\Big(
1+\cos\frac{\chi}{4}(\tfrac{|\vec{b}_{T}|}{1-x})
+ \frac{\chi}{8}(\tfrac{|\vec{b}_{T}|}{1-x})
\Big[\frac{\chi}{4}(\tfrac{|\vec{b}_{T}|}{1-x})-
  \sin\frac{\chi}{4}(\tfrac{|\vec{b}_{T}|}{1-x})
\Big]\Big).
\label{eq:eikonalLensFunctionbTna}
\end{equation}
In principle, SU(3) color can be implemented numerically into the
formalism, however it is non-trivial to derive an analytical expression
similar to Eqs.~(\ref{eq:eikonalLensFunctionTa}) and (\ref{eq:eikonalLensFunctionbTna}). 
The eikonal phase in (\ref{eq:Phase}) 
contains two characteristic quantities of this formalism,
the coupling $\alpha$ and the gauge propagator $\mathcal{D}^{-1}$.
They represent the exchanged soft gluons between quark and spectator.
In order to numerically estimate the lensing function
and in turn the Sivers function 
we need to implement the infrared behavior of the gluon
and the running coupling in the non-perturbative regime where we
  infer
that the soft gluon transverse momentum defines the scale at which
the coupling is evaluated. 
These two quantities 
have been extensively studied in the infrared limit in the
 Dyson-Schwinger framework
\cite{Alkofer:2008tt}
and in lattice QCD \cite{Sternbeck:2008mv}. 
We use calculations of these quantities from  Dyson-Schwinger equations~\cite{Alkofer:2008tt}
where both $\alpha_s$ and $\mathcal{D}^{-1}$ 
are defined in the infrared limit.
With these inputs we plot the lensing function 
versus $|\vec{b}_T|$ for the perturbative, U(1), and  SU(2) cases 
in  Fig.~\ref{fig:Lens-GPD-Sivers}.  We find that for all cases
the lensing function is ``attractive'' as suggested in the general
arguments of Burkardt~\cite{Burkardt:2003uw}.
In order to study the dependence of the relation (\ref{eq:Relation})
on the GPD $E$, we compare the diquark model result, Eq.~(\ref{eq:E})
to another phenomenological model investigated in~\cite{Diehl:2004cx}. 
In Fig.~\ref{fig:Lens-GPD-Sivers} we plot the first moment of
the Sivers function for the various cases which range from $0.04-0.08$. This
is in reasonable agreement with 
the extractions from the Torino and Bochum
groups is $\approx 0.05$~\cite{Anselmino:2008sga,Arnold:2008ap}. 
In this calculation 
we have shown that Eq.~(\ref{eq:Relation}) is indeed preserved when
higher order diagrams contribute to the lensing function 
under conditions of the eikonal and ladder approximation. 
We have also provided an improvement to the 
one-gluon exchange approximation in calculating
T-odd TMDs in the spectator framework.

\begin{theacknowledgments}
L.G. thanks Jacques Soffer for the opportunity to present this work at CIPANP 09. We also thank Alessandro Bacchetta, Stan Brodsky, Matthias Burkardt, Gary Goldstein, Simonetta Liuti and Andreas Metz for useful discussions. 
L.G. acknowledges support from  U.S. Department of Energy under 
contract DE-FG02-07ER41460.  Notice: Authored by Jefferson Science 
Associates, LLC under U.S. DOE Contract No. DE-AC05-06OR23177. The U.S. Government retains a non-exclusive, paid-up, irrevocable, world-wide license to publish or reproduce this manuscript for U.S. Government purposes. 

\end{theacknowledgments}

\bibliographystyle{aipproc}   

\bibliography{cip}

\begin{thebibliography}{30}
\expandafter\ifx\csname natexlab\endcsname\relax\def\natexlab#1{#1}\fi
\providecommand{\enquote}[1]{``#1''}
\expandafter\ifx\csname url\endcsname\relax
  \def\url#1{\texttt{#1}}\fi
\expandafter\ifx\csname urlprefix\endcsname\relax\def\urlprefix{URL }\fi
\providecommand{\eprint}[2][]{\url{#2}}

\bibitem[Sivers(1990)]{Sivers:1989cc}
D.~W. Sivers, \emph{Phys. Rev.} \textbf{D41}, 83 (1990).

\bibitem[Boer and Mulders(1998)]{Boer:1997nt}
D.~Boer, and P.~J. Mulders, \emph{Phys. Rev.} \textbf{D57}, 5780--5786 (1998),
  \eprint{hep-ph/9711485}.

\bibitem[Mulders and Tangerman(1996)]{Mulders:1995dh}
P.~J. Mulders, and R.~D. Tangerman, \emph{Nucl. Phys.} \textbf{B461}, 197--237
  (1996), \eprint{hep-ph/9510301}.

\bibitem[Ji et~al.(2005)]{Ji:2004wu}
X.-d. Ji, J.-p. Ma, and F.~Yuan, \emph{Phys. Rev.} \textbf{D71}, 034005 (2005),
  \eprint{hep-ph/0404183}.

\bibitem[Collins(2002)]{Collins:2002kn}
J.~C. Collins, \emph{Phys. Lett.} \textbf{B536}, 43--48 (2002),
  \eprint{hep-ph/0204004}.

\bibitem[Belitsky et~al.(2003)]{Belitsky:2002sm}
A.~V. Belitsky, X.~Ji, and F.~Yuan, \emph{Nucl. Phys.} \textbf{B656}, 165--198
  (2003), \eprint{hep-ph/0208038}.

\bibitem[Boer et~al.(2003)]{Boer:2003cm}
D.~Boer, P.~J. Mulders, and F.~Pijlman, \emph{Nucl. Phys.} \textbf{B667},
  201--241 (2003), \eprint{hep-ph/0303034}.

\bibitem[Brodsky et~al.(2002)]{Brodsky:2002cx}
S.~J. Brodsky, D.~S. Hwang, and I.~Schmidt, \emph{Phys. Lett.} \textbf{B530},
  99--107 (2002), \eprint{hep-ph/0201296}.

\bibitem[Ji and Yuan(2002)]{Ji:2002aa}
X.-d. Ji, and F.~Yuan, \emph{Phys. Lett.} \textbf{B543}, 66--72 (2002),
  \eprint{hep-ph/0206057}.

\bibitem[Goldstein and Gamberg(2002)]{Goldstein:2002vv}
G.~R. Goldstein, and L.~Gamberg  (2002), published in Amsterdam ICHEP 452-454,
  \eprint{hep-ph/0209085}.

\bibitem[Gamberg et~al.(2003)]{Gamberg:2003ey}
L.~P. Gamberg, G.~R. Goldstein, and K.~A. Oganessyan, \emph{Phys. Rev.}
  \textbf{D67}, 071504 (2003), \eprint{hep-ph/0301018}.

\bibitem[Bacchetta et~al.(2004)]{Bacchetta:2003rz}
A.~Bacchetta, A.~Schaefer, and J.-J. Yang, \emph{Phys. Lett.} \textbf{B578},
  109--118 (2004), \eprint{hep-ph/0309246}.

\bibitem[Lu and Ma(2004)]{Lu:2004au}
Z.~Lu, and B.-Q. Ma, \emph{Nucl. Phys.} \textbf{A741}, 200--214 (2004),
  \eprint{hep-ph/0406171}.

\bibitem[Gamberg et~al.(2008)]{Gamberg:2007wm}
L.~P. Gamberg, G.~R. Goldstein, and M.~Schlegel, \emph{Phys. Rev.}
  \textbf{D77}, 094016 (2008), \eprint{0708.0324}.

\bibitem[Bacchetta et~al.(2008)]{Bacchetta:2008af}
A.~Bacchetta, F.~Conti, and M.~Radici, \emph{Phys. Rev.} \textbf{D78}, 074010
  (2008), \eprint{0807.0323}.

\bibitem[Gamberg and Schlegel(2009)]{gambschl:2009gs}
L.~Gamberg, and M.~Schlegel  (2009), in preparation.

\bibitem[Abarbanel and Itzykson(1969)]{Abarbanel:1969ek}
H.~D.~I. Abarbanel, and C.~Itzykson, \emph{Phys. Rev. Lett.} \textbf{23}, 53
  (1969).

\bibitem[Fried et~al.(2000)]{Fried:2000hj}
H.~M. Fried, Y.~Gabellini, and J.~Avan, \emph{Eur. Phys. J.} \textbf{C13},
  699--709 (2000).

\bibitem[Burkardt(2002)]{Burkardt:2002ks}
M.~Burkardt, \emph{Phys. Rev.} \textbf{D66}, 114005 (2002),
  \eprint{hep-ph/0209179}.

\bibitem[Burkardt(2004)]{Burkardt:2003uw}
M.~Burkardt, \emph{Nucl. Phys.} \textbf{A735}, 185--199 (2004),
  \eprint{hep-ph/0302144}.

\bibitem[Meissner et~al.(2007)]{Meissner:2007rx}
S.~Meissner, A.~Metz, and K.~Goeke, \emph{Phys. Rev.} \textbf{D76}, 034002
  (2007), \eprint{hep-ph/0703176}.

\bibitem[Meissner et~al.(2008)]{Meissner:2008ay}
S.~Meissner, A.~Metz, M.~Schlegel, and K.~Goeke, \emph{JHEP} \textbf{08}, 038
  (2008), \eprint{0805.3165}.

\bibitem[Meissner et~al.(2009)]{Meissner:2009ww}
S.~Meissner, A.~Metz, and M.~Schlegel  (2009), \eprint{0906.5323}.

\bibitem[Burkardt and Hwang(2004)]{Burkardt:2003je}
M.~Burkardt, and D.~S. Hwang, \emph{Phys. Rev.} \textbf{D69}, 074032 (2004),
  \eprint{hep-ph/0309072}.

\bibitem[Gamberg et~al.(2006)]{Gamberg:2006ru}
L.~P. Gamberg, D.~S. Hwang, A.~Metz, and M.~Schlegel, \emph{Phys. Lett.}
  \textbf{B639}, 508--512 (2006), \eprint{hep-ph/0604022}.

\bibitem[Alkofer et~al.(2009)]{Alkofer:2008tt}
R.~Alkofer, C.~S. Fischer, F.~J. Llanes-Estrada, and K.~Schwenzer, \emph{Annals
  Phys.} \textbf{324}, 106--172 (2009), \eprint{0804.3042}.

\bibitem[Sternbeck and von Smekal(2008)]{Sternbeck:2008mv}
A.~Sternbeck, and L.~von Smekal  (2008), \eprint{0811.4300}.

\bibitem[Diehl et~al.(2005)]{Diehl:2004cx}
M.~Diehl, T.~Feldmann, R.~Jakob, and P.~Kroll, \emph{Eur. Phys. J.}
  \textbf{C39}, 1--39 (2005), \eprint{hep-ph/0408173}.

\bibitem[Anselmino et~al.(2009)]{Anselmino:2008sga}
M.~Anselmino, et~al., \emph{Eur. Phys. J.} \textbf{A39}, 89--100 (2009),
  \eprint{0805.2677}.

\bibitem[Arnold et~al.(2008)]{Arnold:2008ap}
S.~Arnold, A.~V. Efremov, K.~Goeke, M.~Schlegel, and P.~Schweitzer  (2008),
  \eprint{0805.2137}.

\end{thebibliography}

\end{document}